\documentclass[sigconf]{acmart}

\usepackage{wrapfig}
\usepackage{multirow}
\usepackage{longtable}
\usepackage{xspace}
\usepackage{tabularray}
\usepackage{tabularx}
\usepackage{enumitem}
\usepackage{caption}
\usepackage{subcaption}
\usepackage{graphicx}

\AtBeginDocument{%
  }

\acmConference[GenAICHI 2025]{Generative AI and HCI: A CHI 2025 Workshop}{April 27, 2025}{Yokohama, Japan and Online}
\acmDOI{}           
\acmISBN{}          

\usepackage{rotating}
\usepackage{tabularx}
\usepackage{graphicx}  
\usepackage{tikz}      
\usepackage{xcolor} 

\begin{document}

\title[LACE]{LACE: Controlled Image Prompting and Iterative Refinement with GenAI for Professional Visual Art Creators} 

\author{Yen-Kai Huang}
\affiliation{%
  \institution{Dartmouth College}
  \city{Hanover}
  \state{New Hampshire}
  \country{USA}
}
\email{yenkai.huang.gr@dartmouth.edu}

\author{Zheng Ning}
\affiliation{%
  \institution{University of Notre Dame}
  \city{Notre Dame}
  \state{Indiana}
  \country{USA}
}
\email{zning@nd.edu}






\keywords{creativity support, generative AI, professional creative tools, image creation}

\begin{teaserfigure}
    \centering
    \includegraphics[width=0.9\linewidth]{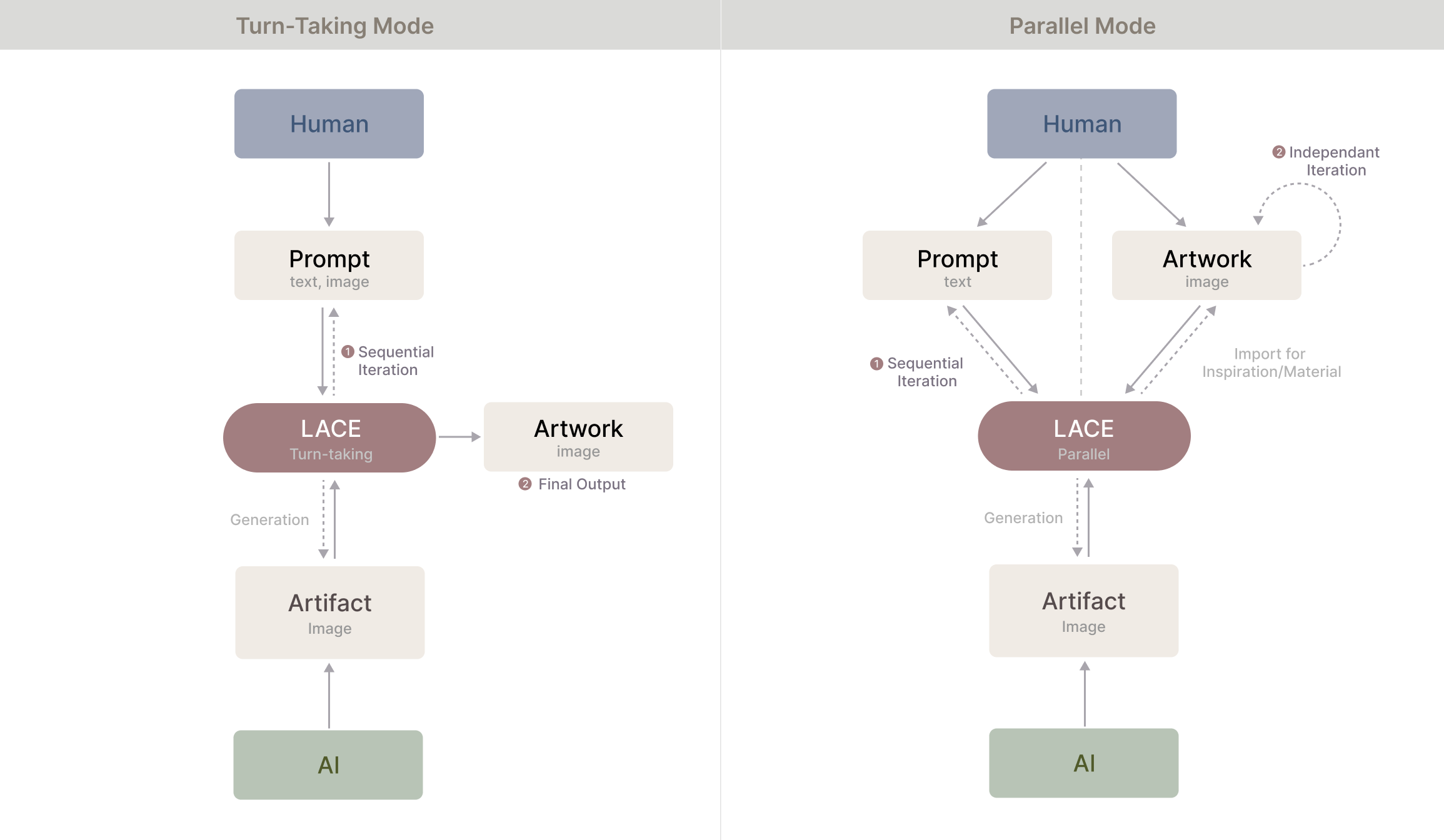}
    \caption{Turn-Taking vs. Parallel: Turn-Taking follows a sequential workflow where human and AI iterate on prompts and outputs step by step. Parallel mode enables simultaneous refinement, with AI adapting to snapshots of the artist’s evolving canvas, fostering a dynamic interplay between human and machine contributions.}
    \label{fig:LACE-mode}
\end{teaserfigure}

\maketitle

\section{INTRODUCTION}

The advent of generative AI has opened exciting possibilities in professional art creation by generating high-quality outputs almost instantly. However, recent research~\cite{Draxler2023ghostWriter,First-Authorship,Workflows_genAI,author2024worriedAI} highlights several challenges that hinder its practical adoption by professional users:

\begin{itemize}
    \item \textbf{Inadequate expressiveness of text-based prompting}. Purely textual descriptions often lack the precision required to control nuanced elements such as lighting, texture patterns, subtle positioning, and overall artistic ambiance. This ambiguity makes it challenging for artists to achieve their intended vision.
    \item \textbf{Lack of coherence in iterative refinement}. Generative workflows are highly sensitive to input parameters~\cite{wu2023uncovering}, making it difficult for artists to progressively build on previous versions and maintain creative continuity.
    \item \textbf{Incompatibility with established workflows}. Many artists are reluctant to switch to unfamiliar AI tools, especially when these tools disrupt their current routines or require specialized technical knowledge~\cite{Workflows_genAI}.
\end{itemize}

To address these issues, we introduce the \emph{Latent Auto-recursive Composition Engine} (LACE), a system that integrates generative AI into industry-standard environments. LACE provides fine-grained control through Photoshop’s familiar editing features (e.g., blending modes, curves, layer-based adjustments), letting artists seamlessly refine and compose AI-generated outputs. By coupling direct manipulation with a customizable AI pipeline, LACE allows users to balance manual control and automated generation, enhancing both creative freedom and iterative coherence.

Our preliminary pilot study ($N = 21$) indicates that LACE significantly improves usability, user ownership, and overall satisfaction compared to baseline AI workflows. In the following sections, we describe LACE’s design, present key findings from our user study, and discuss implications for integrating generative AI into professional art practices. 

\begin{figure}
    \centering
    \begin{subfigure}[b]{0.49\textwidth}
        \centering
        \begin{tikzpicture}
            \node[anchor=south west,inner sep=0] (image) at (0,0) {\includegraphics[width=\textwidth]{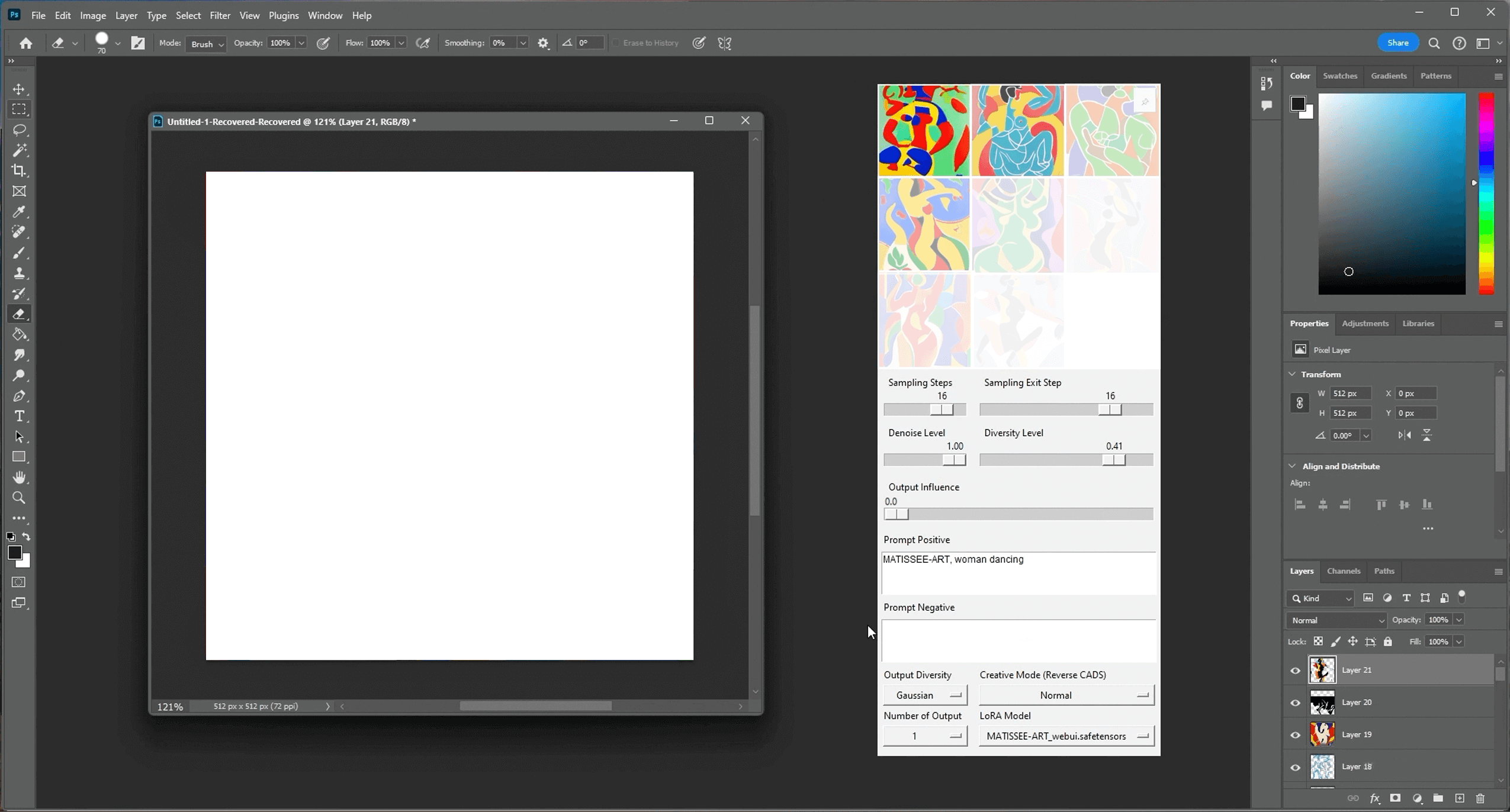}};
        \end{tikzpicture}
        \caption{LACE integrates into Photoshop via a popup GUI, linking manual edits with AI.}
        \label{fig:subfig1}
    \end{subfigure}
    \hfill
    \begin{subfigure}[b]{0.49\textwidth}
        \centering
        \begin{tikzpicture}
            \node[anchor=south west,inner sep=0] (image) at (0,0) {\includegraphics[width=\textwidth]{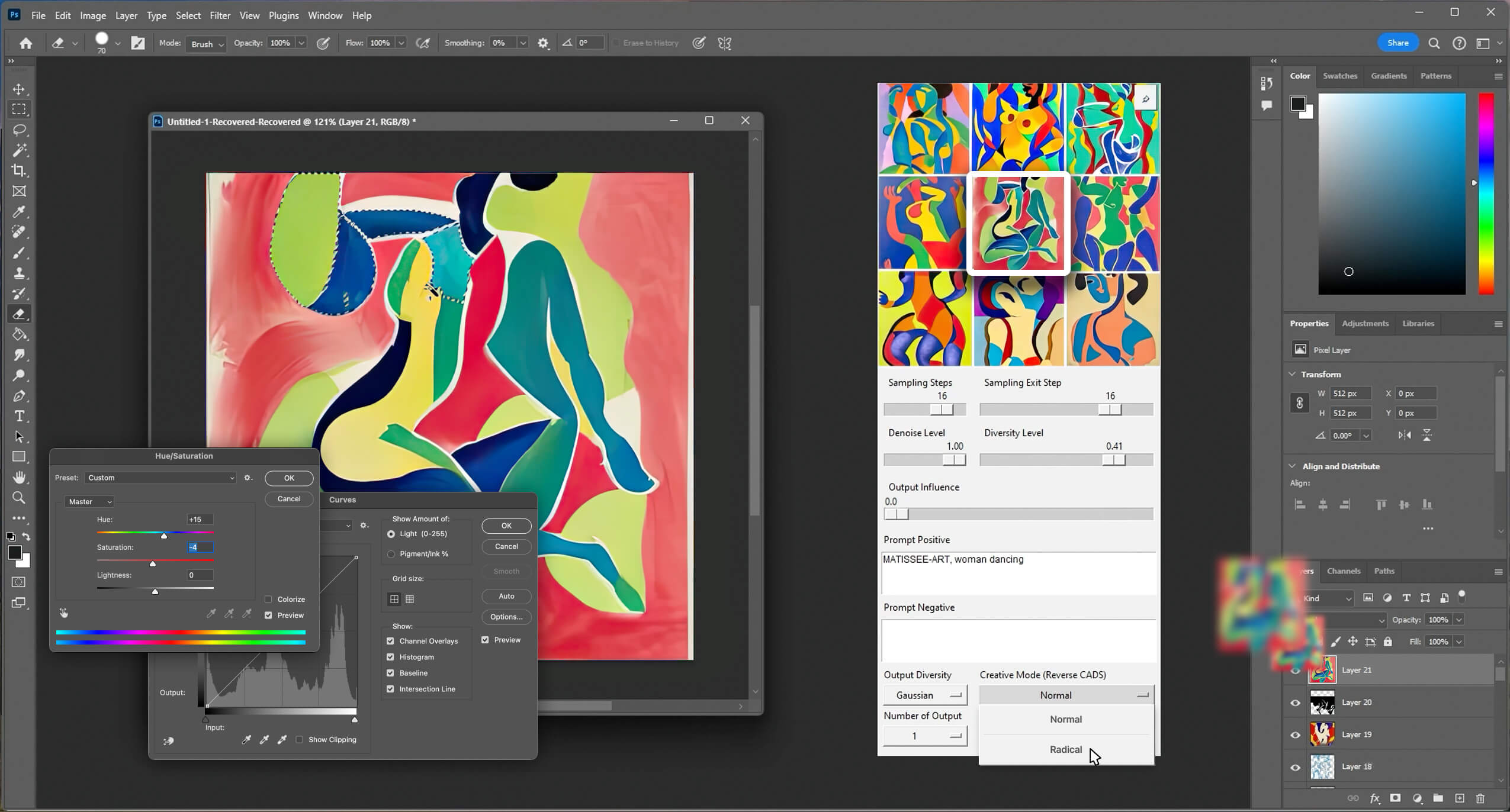}};
        \end{tikzpicture}
        \caption{LACE uses Photoshop’s robust tools to guide precise image prompting.}
        \label{fig:subfig2}
    \end{subfigure}
    \vfill
    \hfill
    \begin{subfigure}[b]{0.49\textwidth}
        \centering
        \begin{tikzpicture}
            \node[anchor=south west,inner sep=0] (image) at (0,0) {\includegraphics[width=\textwidth]{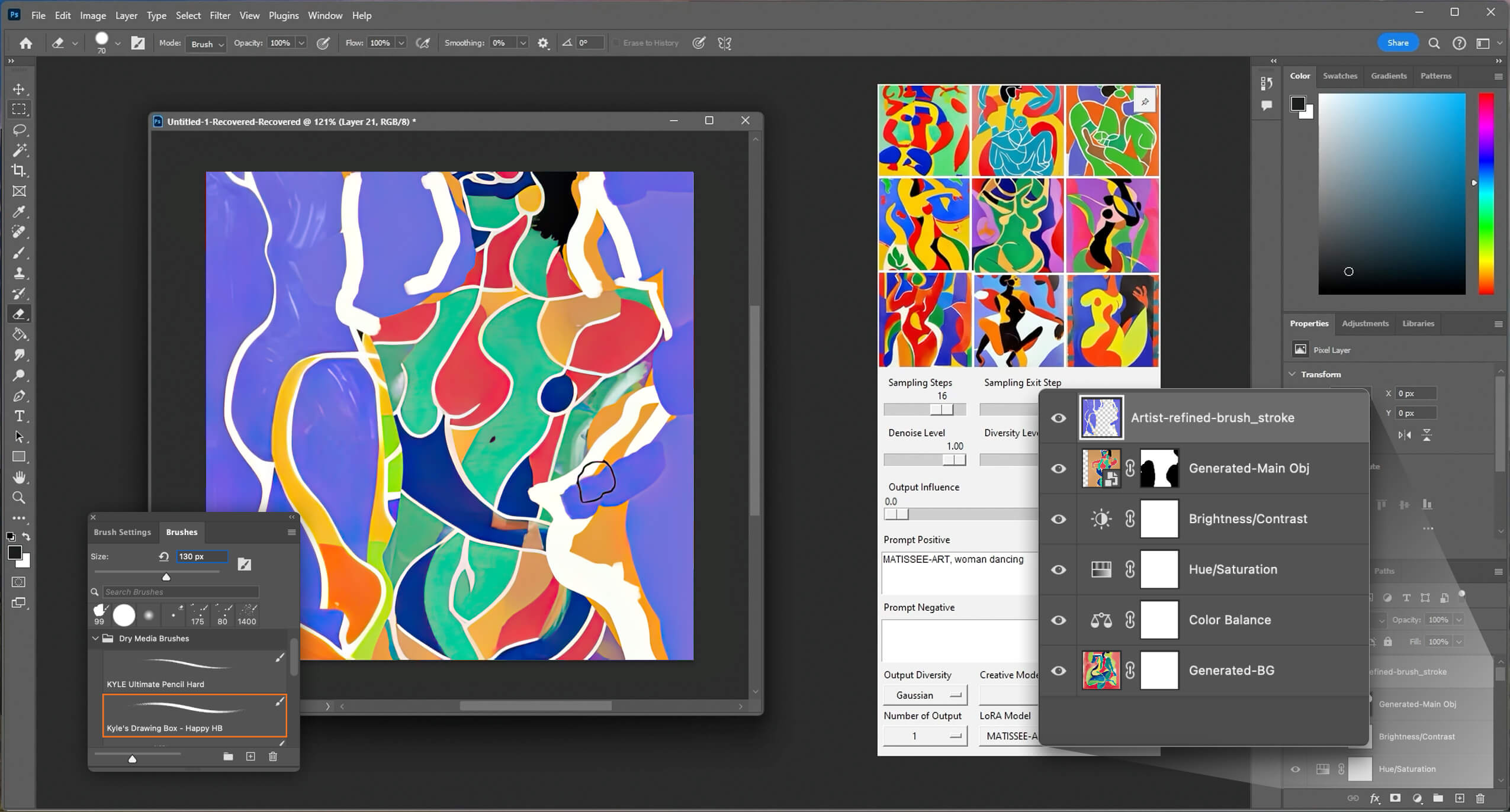}};
        \end{tikzpicture}
        \caption{By layering AI outputs in Photoshop, LACE keeps artwork consistent without extra tuning.}
        \label{fig:subfig3}
    \end{subfigure}
    \hfill
    \begin{subfigure}[b]{0.49\textwidth}
        \centering
        \begin{tikzpicture}
            \node[anchor=south west,inner sep=0] (image) at (0,0) {\includegraphics[width=\textwidth]{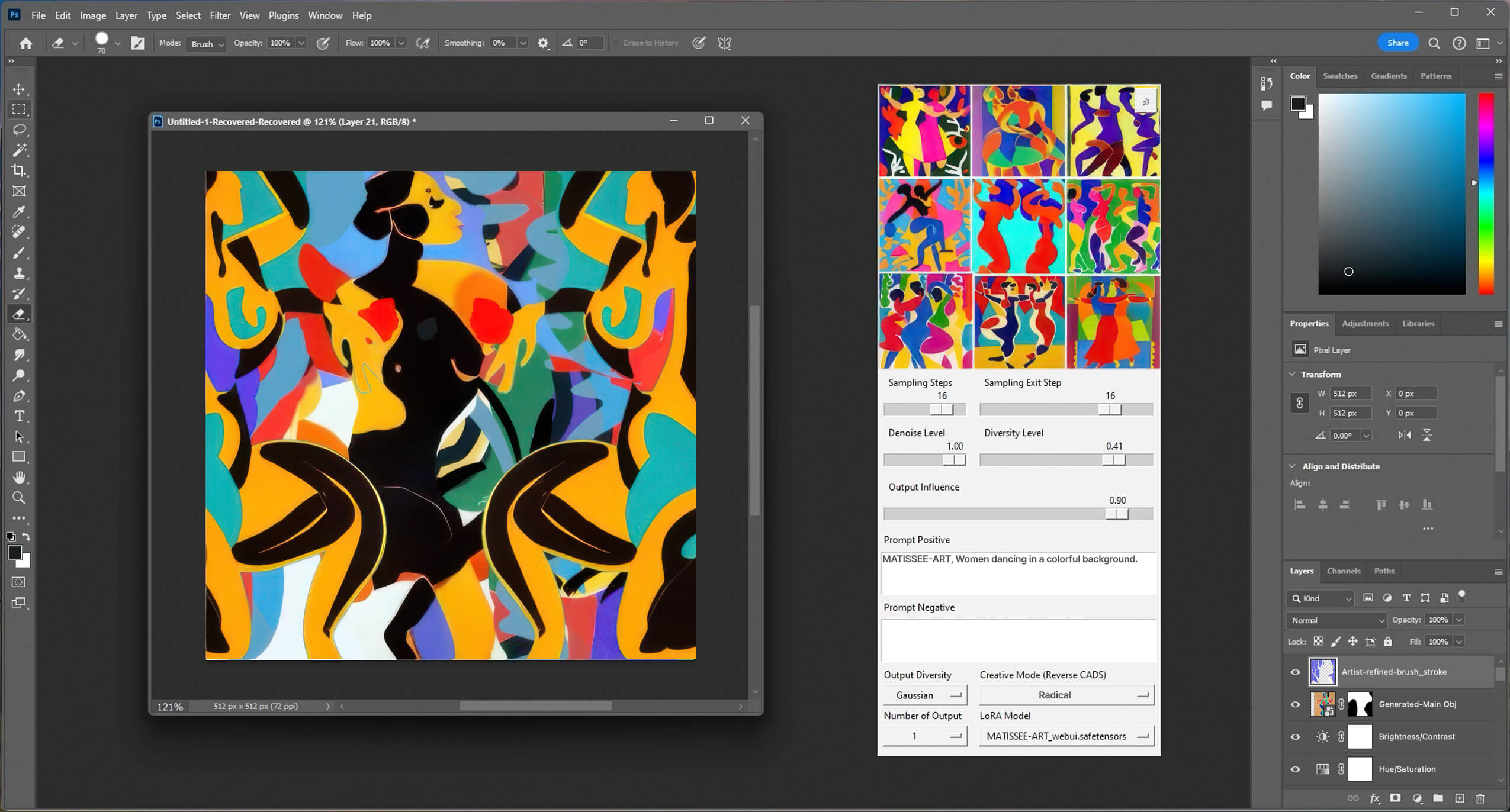}};
        \end{tikzpicture}
        \caption{AI adapts in real time to the evolving canvas, creating a dynamic two-way creative exchange.}
        \label{fig:subfig4}
    \end{subfigure}
    \caption{Overview of LACE: Interface and Interaction}
    \label{fig:combined}
\end{figure}

\section{LACE SYSTEM}
LACE bridges Photoshop and an AI pipeline (Fig~\ref{fig:LACE-arc}), allowing users to edit AI-generated elements as individual layers rather than as flattened images. By leveraging Photoshop’s native features (e.g., layers, masks, blending modes, and color adjustments), LACE maintains familiar workflows while integrating AI-assisted image creation~\cite{semantic_segmentation_diffusion, cao2023animediffusion} and generative control mechanisms~\cite{feng2024item, yang2024fine}. Grounded in the COFI framework~\cite{COFI} by Rezwana et al., LACE supports both turn-taking and parallel collaboration for flexible co-creative interactions. Additionally, LACE’s modular design allows users to configure or replace the underlying AI pipeline and connect new models directly to Photoshop.


\subsection{Layer-Based Prompting} 
\begin{figure}[h]
    \centering
    \begin{minipage}{1\linewidth}
        \centering
        \includegraphics[width=\linewidth]{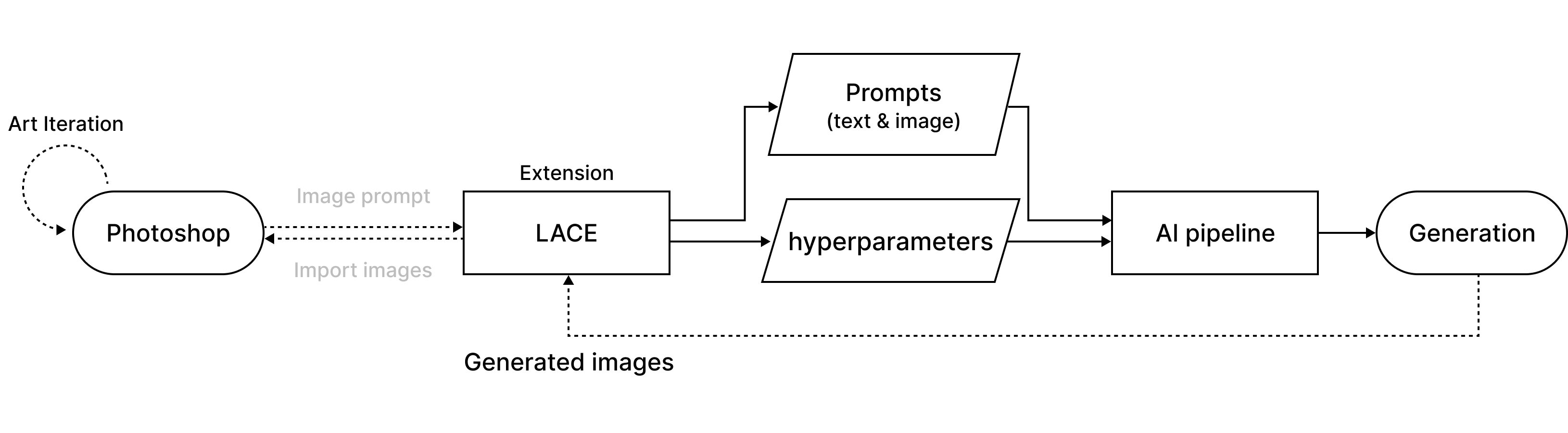}
        \caption{LACE’s modular design lets users configure or replace the underlying AI pipeline.}
        \label{fig:LACE-arc}
    \end{minipage}
    \hfill
    \begin{minipage}{1\linewidth}
        \centering
        \includegraphics[width=\linewidth]{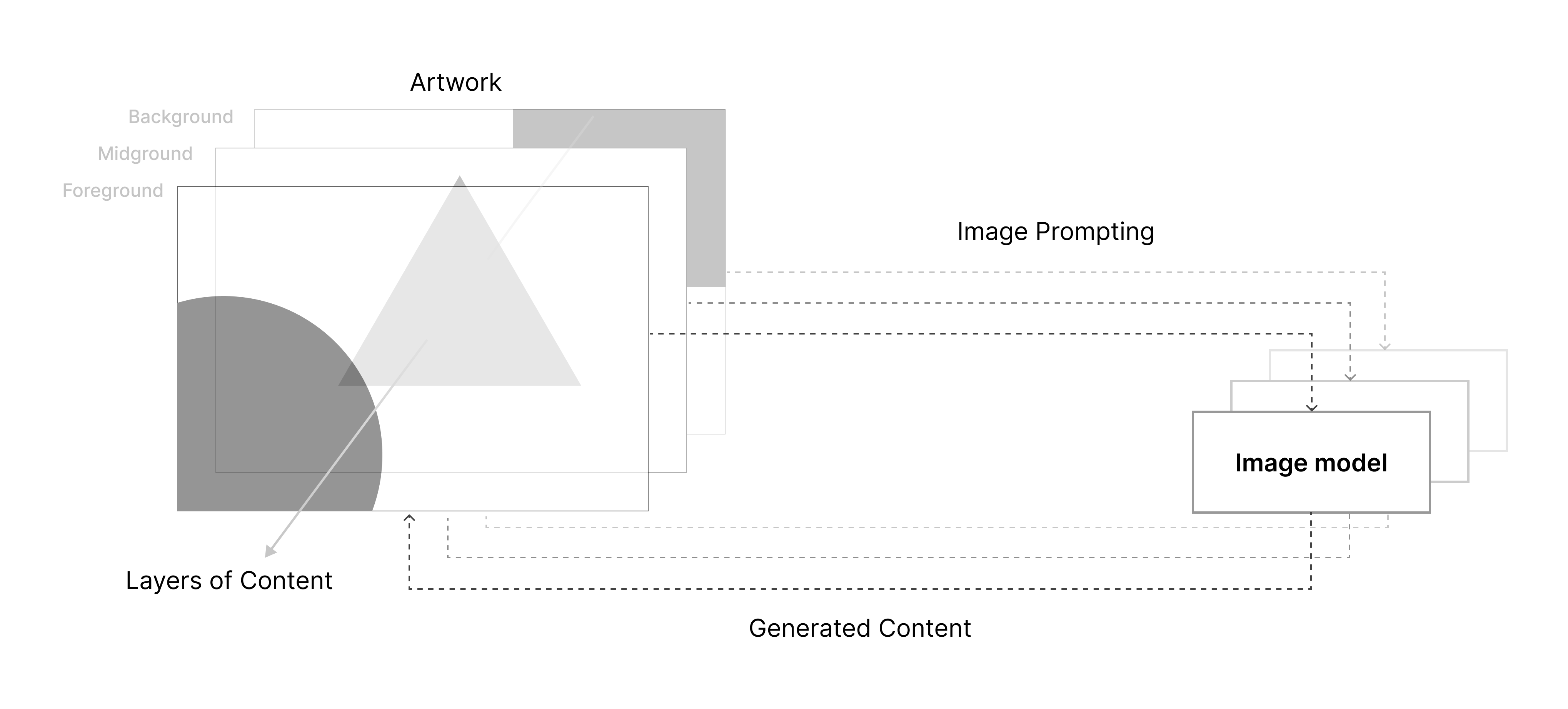}
        \caption{Layer-based prompting offers granular control over AI-generated elements.}
        \label{fig:LACE-layer}
    \end{minipage}
\end{figure}

Instead of generating flattened images, LACE imports AI outputs into Photoshop as separate layers (Fig~\ref{fig:LACE-layer}), enabling creators to isolate and modify foreground, mid-ground, and background elements independently without retraining or fine-tuning models. Conventional AI tools produce a single, flattened image, making it challenging to adjust individual components without extensive manual editing. In contrast, LACE allows artists to create distinct layers for different parts of their artwork, offering a level of control and precision that traditional methods cannot achieve. This layer-based approach ensures that the iterative refinement process remains flexible and artist-driven while continuously integrating AI-generated content.

\subsection{Flexible Collaboration Modes}
LACE integrates with Photoshop’s editing timeline, enabling continuous interaction between the artist and AI (Fig~\ref{fig:LACE-mode}). In this workflow, artists refine images while the AI generates suggestions, forming a feedback loop that maintains creative continuity. This structure supports both sequential turn-taking for precise adjustments and parallel interaction, where AI suggestions evolve as the artist works. Fluid transitions between these modes enhance flexibility, allowing creators to tailor AI collaboration to their process.
    \begin{itemize}
        \item \textbf{Turn-Taking mode}: Starting from scratch, users provide text prompts that generate AI outputs, forming the basis for iterative, step-by-step refinement.
        \item \textbf{Parallel mode}: AI adapts to the evolving canvas, producing outputs that inspire the artist while incorporating their refinements in real time.
    \end{itemize}

\section{STUDY AND KEY FINDINGS}

We conducted a within-subject pilot study with 21 participants (ages 20–27; 45\% beginners, 25\% intermediate, 30\% advanced) to assess LACE’s impact on creativity and workflow. They completed one of three art tasks (Representational, Non-Representational, or Design Challenge) using three workflows: W1 (text-to-image), W2 (text-to-image with latent consistency), and W3 (LACE, image-to-image with latent consistency). Workflow order was randomized, with each session lasting 15 minutes. All generative model parameters remained consistent across conditions.

\subsection{Key Findings}

Quantitative findings show a strong preference for LACE (W3) over the two text-based workflows (W1 and W2). A Friedman test identified significant differences in satisfaction ($p = 0.039$), ownership ($p = 0.009$), usability ($p = 0.003$), and artistic perception ($p = 0.005$). Post-hoc comparisons indicate that LACE outperforms both text-only (W1) and text-based iterative (W2) approaches, most notably in usability and the sense of ownership.

\subsubsection{\textbf{Improved Usability and Ownership}} 
Participants found it easier to position and refine image elements using LACE’s layer-based editing compared to purely text-based methods. They reported higher artistic control and fewer prompt-driven uncertainties. Overall, 71.4\% selected LACE as their favorite workflow, citing its ability to retain and combine favorable elements from earlier iterations. The pixel-art design challenge (T3) generated mixed feedback; although users appreciated the deeper control, some found it more time-intensive.

\subsubsection{\textbf{Correlation Between Ownership and Usability}} 
Spearman correlation analyses reveal that higher usability ratings in LACE are associated with stronger ownership and satisfaction scores. By contrast, text-based workflows do not show a similar relationship—participants who found the interface straightforward did not consistently feel satisfied with the final outputs. Some noted difficulties in generating precise compositions or styles through text alone, highlighting a gap between prompt adjustments and desired visual outcomes.

\subsubsection{\textbf{Engagement in Structured Tasks}}
Time logs indicate that participants spent more effort with LACE, particularly on structured design tasks like T3. While text-based prompts required fewer steps, LACE’s iterative image edits encouraged ongoing refinements. This suggests that LACE may increase both creative investment and the learning curve for users less familiar with advanced image-editing features. However, participants generally viewed the extra effort as worthwhile, citing improved alignment between their vision and the final results.

\section{DISCUSSION AND FUTURE WORK}

Our findings suggest that the choice of Human-AI collaboration mode depends on both the creative task and its stage. Participants generally preferred turn-taking during early ideation, where iterative text prompts facilitated concept exploration, while later, detail-oriented stages benefited from parallel interaction for real-time refinements. However, these preferences are not rigid; the interplay of timing, task complexity, and user goals remains nuanced and warrants further investigation. Future research will examine how different workflows and phases of the creative process influence user perception and behavior.

Interestingly, participants favored LACE even when its outputs were suboptimal, suggesting that artistic agency during iteration significantly impacts user experience. This highlights the importance of control and interactivity in AI-assisted creativity. Moving forward, our goal is to extend LACE’s design principles to other creative domains supported by AI, such as 3D rendering, animation, and graphic design, where fine-grained control and iterative workflows are essential. Future work will also explore adapting collaboration modes for different skill levels and optimizing layer-based editing to enhance usability without overwhelming novice users.

\bibliographystyle{ACM-Reference-Format}
\bibliography{ref}
\balance
\clearpage

\onecolumn
\appendix
\section{SAMPLED QUALITATIVE RESULTS}

\begin{figure}[!ht]
    \centering
    \includegraphics[width=.6\columnwidth]{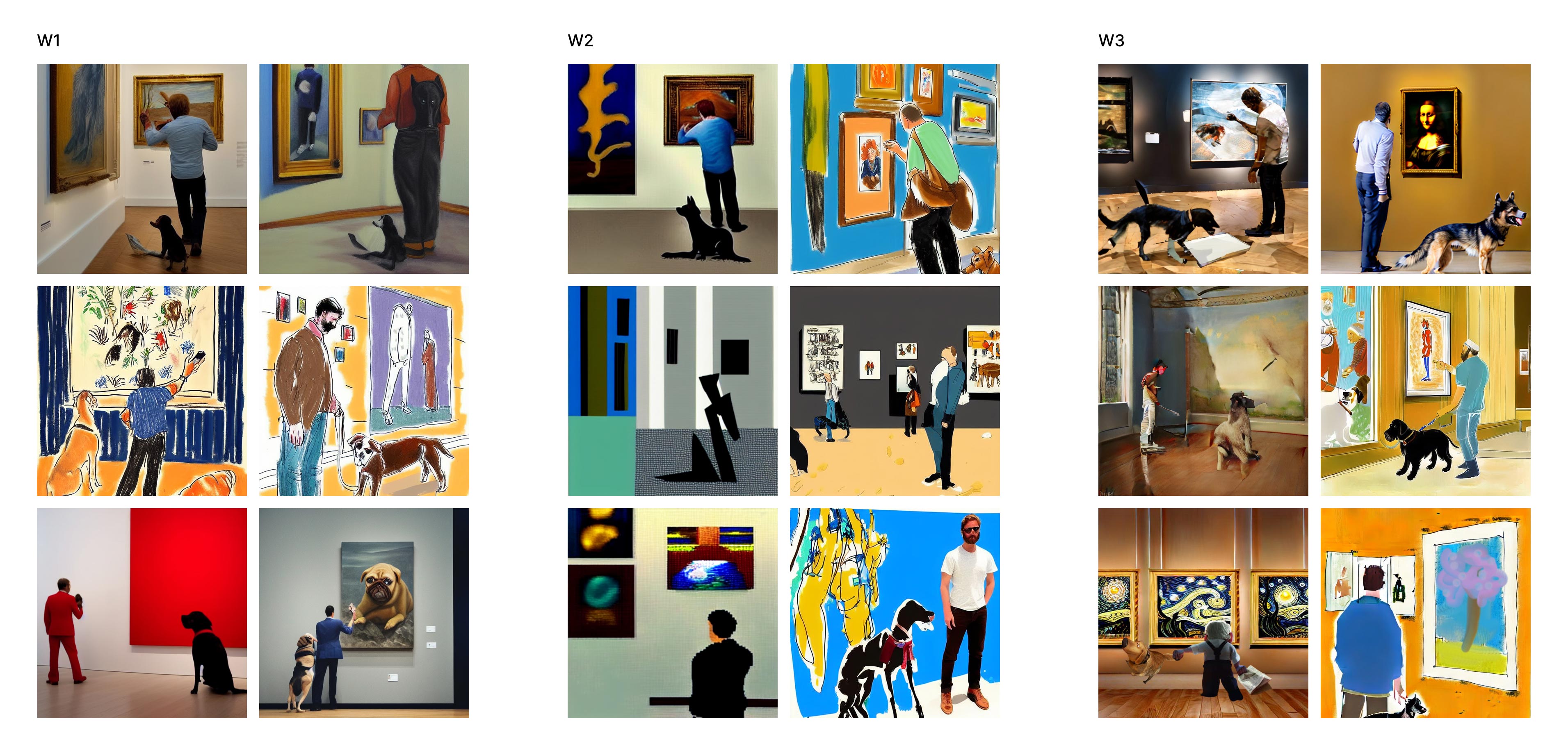}
    \caption{Results from Task 1. 
    Prompt: \texttt{``A man reaching for a painting...''}}
    \label{fig:result_T1}
\end{figure}

\begin{figure}[!ht]
    \centering
    \includegraphics[width=.6\columnwidth]{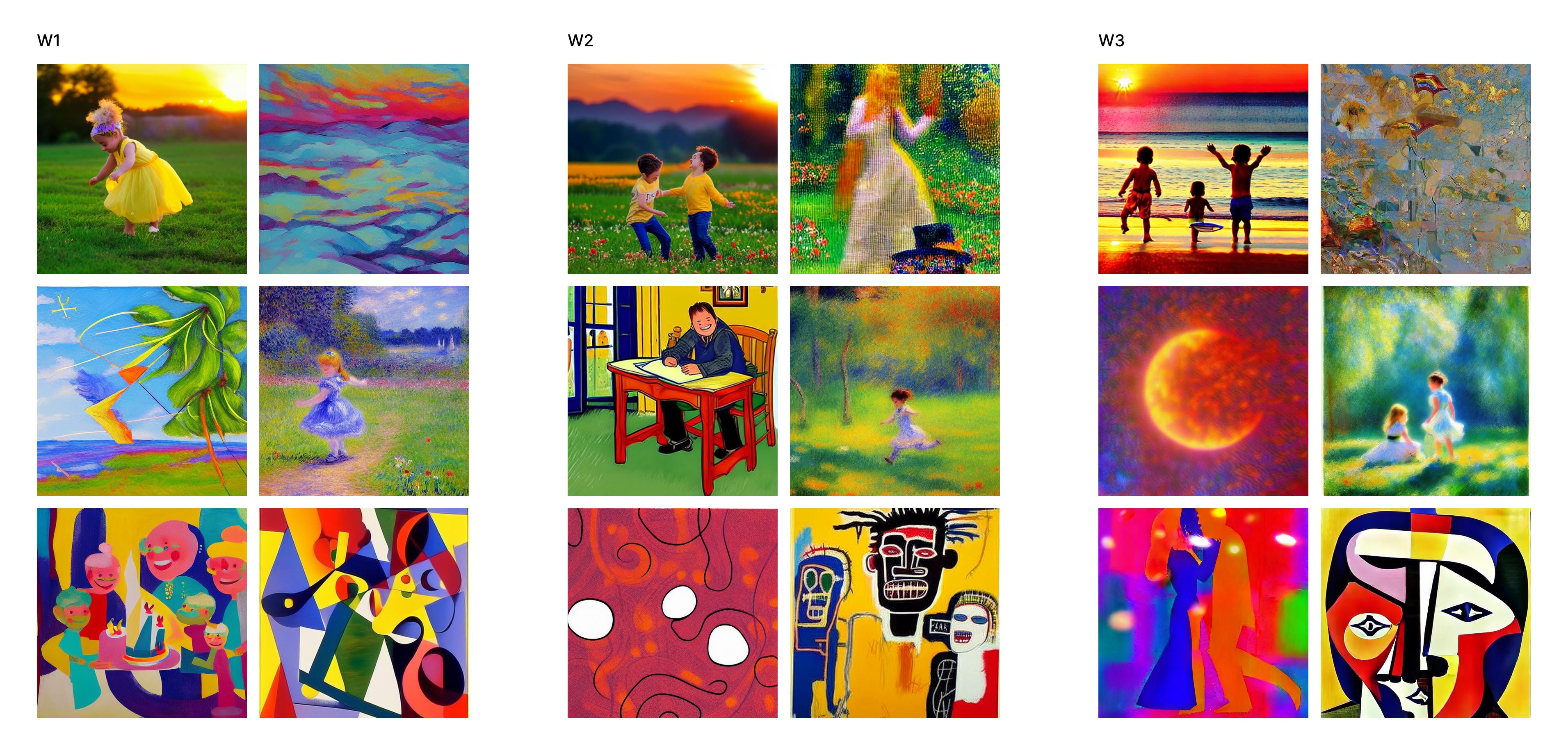}
    \caption{Results from Task 2. 
    Prompt: \texttt{``An abstract composition that embodies...''}}
    \label{fig:result_T2}
\end{figure}

\begin{figure}[!ht]
    \centering
    \includegraphics[width=.6\columnwidth]{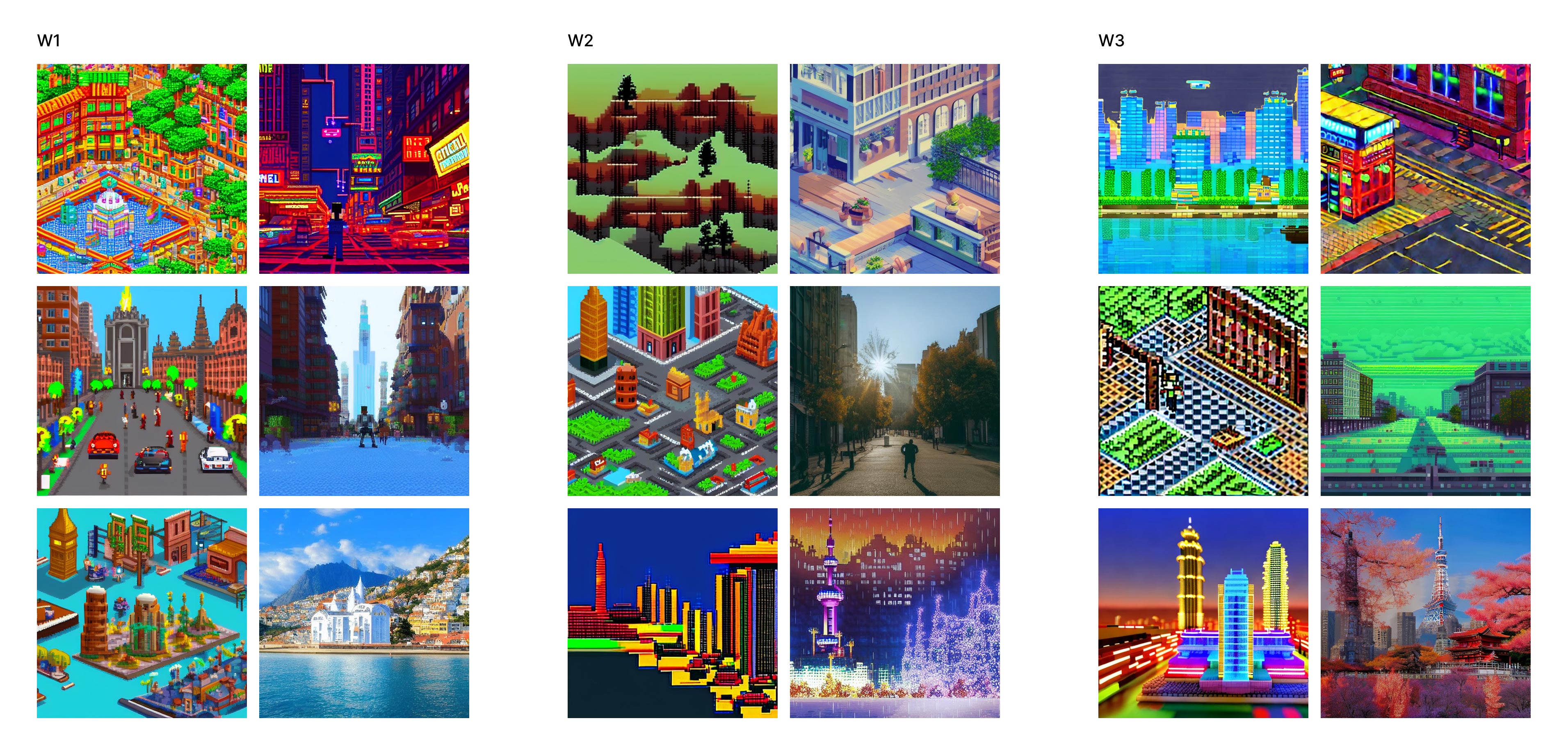}
    \caption{Results from Task 3. 
    Prompt: \texttt{``A pixel art game scene...''}}
    \label{fig:result_T3}
\end{figure}

\begin{figure}[!ht]
    \centering
    \includegraphics[width=.8\columnwidth]{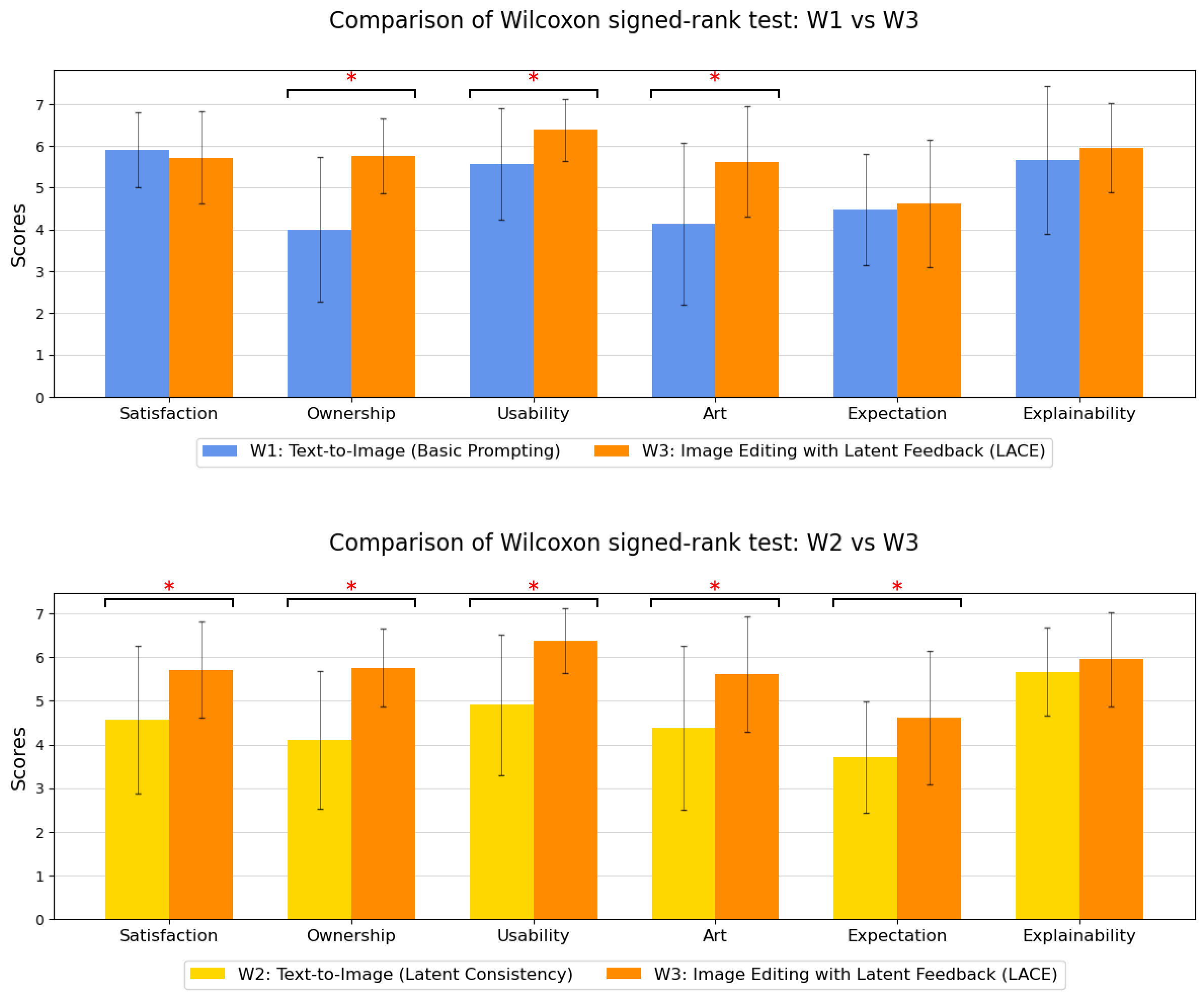}
    \caption{The charts compare Workflow 1 and Workflow 2 
    against Workflow 3 (LACE) using Likert scale scores (1-7)...}
    \label{fig:Likert-overall}
    \vspace{5em} 
\end{figure}

\begin{figure}[!ht]
    \centering
    \includegraphics[width=\columnwidth]{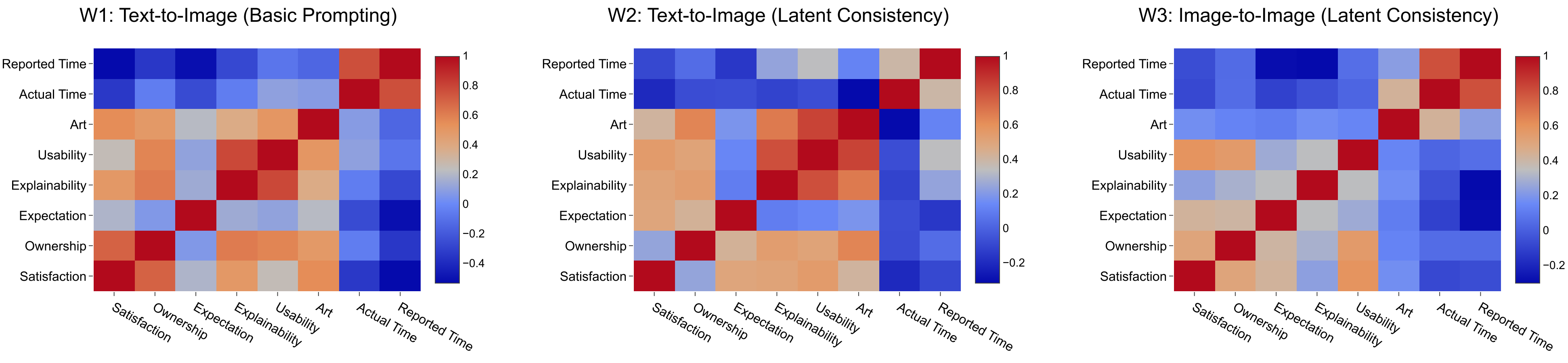}
    \caption{Spearman Correlation Heatmap of Likert-Scale 
    Variables and task completion time}
    \label{fig:corr-workflow}
\end{figure}

\begin{figure}[!ht]
    \centering
    \includegraphics[width=\columnwidth]{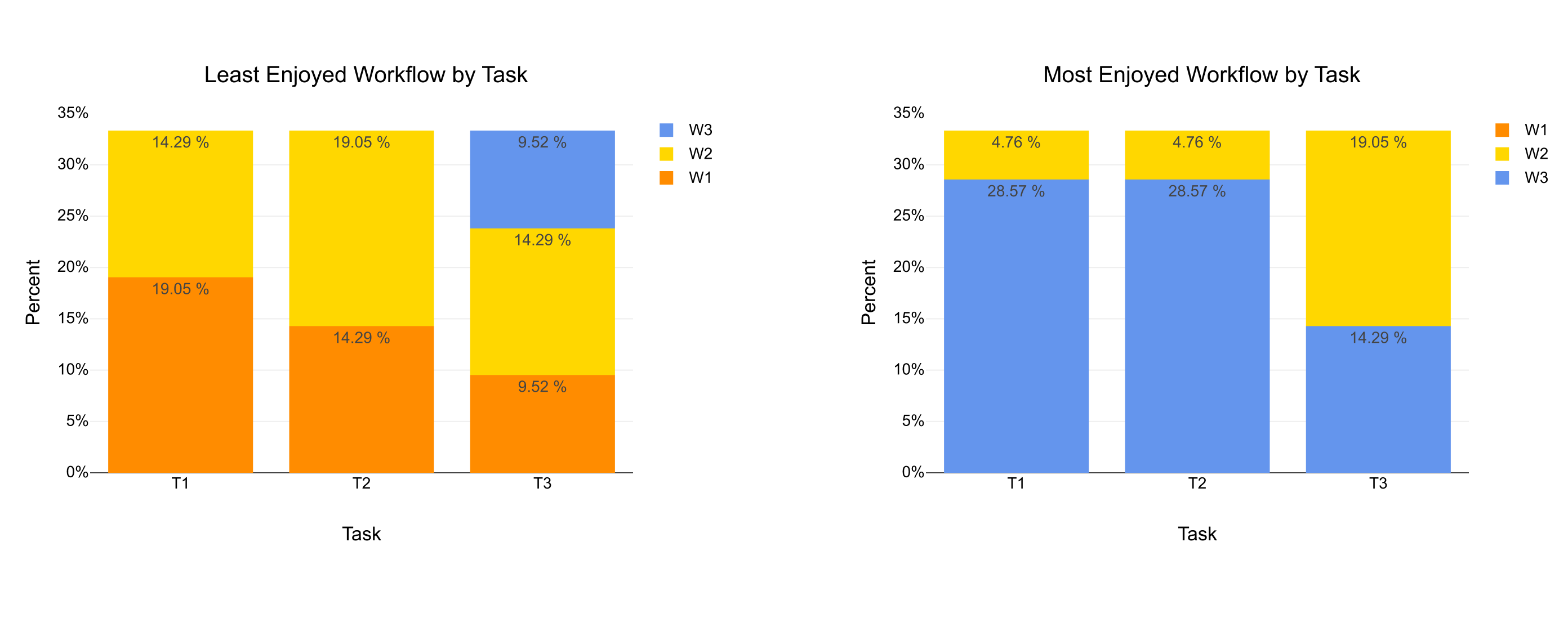}
    \caption{Preference of Workflow by Tasks}
    \label{fig:enjoy-w}
\end{figure}

\end{document}